\documentclass[12pt]{article}
\usepackage{amssymb}
\usepackage{cite} 
\usepackage[english]{babel}
\usepackage{hyperref}
\usepackage{xcolor}
\usepackage{bbold} 

\newcommand{\be}{\begin{equation}}
\newcommand{\ee}{\end{equation}}
\newcommand{\bea}{\begin{array}}
\newcommand{\ea}{\end{array}}
\newcommand{\beqa}{\begin{eqnarray}}
\newcommand{\eeqa}{\end{eqnarray}}
\newcommand{\bean}{\begin{eqnarray*}}
\newcommand{\eean}{\end{eqnarray*}}

\newcommand{\gapproxeq}{\lower
.7ex\hbox{$\;\stackrel{\textstyle >}{\sim}\;$}}
\newcommand{\lapproxeq}{\lower .7ex\hbox{$\;\stackrel
{\textstyle <}{\sim}\;$}}

\newcounter{appendice}

\def\thebibliography#1{{\bf REFERENCES\markboth
{REFERENCES}{REFERENCES}}\list
{[\arabic{enumi}]}{\settowidth\labelwidth{[#1]}\leftmargin\labelwidth
\advance\leftmargin\labelsep
\usecounter{enumi}}
\def\newblock{\hskip .11em plus .33em minus -.07em}
\sloppy
\sfcode`\.=1000\relax}

\def\BI{{\mathbb 1}} 

\def\kbar{{\rlap{$k$}{}^{-}}}

\definecolor{patriarch}{rgb}{0.9, 0.0, 0.5}

\begin{document}

\centerline{\LARGE Asymptotic commutativity of quantized spaces:}
\vskip .25cm
\centerline{\LARGE  the case of  $\mathbb{CP}^{p,q}$ }
\vskip 1cm


\centerline{Fedele Lizzi${}^{1,2,3}$\footnote{fedele.lizzi@na.infn.it}, A. Pinzul${}^4$\footnote{apinzul@unb.br}, A. Stern$^{5}$\footnote{astern@ua.edu} and   Chuang Xu$^{5}\footnote{cxu24@crimson.ua.edu} $ }

\vskip 1cm
\begin{center}
 {${}^1$
Dipartimento di Fisica “Ettore Pancini”, Universit\`a di Napoli
Federico II, Napoli, Italy\\}

 {${}^2$
INFN, Sezione di Napoli, Italy\\}


 {${}^3$ Departament de F\'{\i}sica Qu\`antica i Astrof\'{\i}sica and Institut de C\'{\i}encies
del Cosmos (ICCUB), Universitat de Barcelona, Barcelona, Spain\\}

  {${}^4$ Universidade  de Bras\'{\i}lia, Instituto de F\'{\i}sica\\
70910-900, Bras\'{\i}lia, DF, Brasil\\
and\\
International Center of Physics\\
C.P. 04667, Bras\'{\i}lia, DF, Brazil
\\}

%
%
%
  {${}^5$ Department of Physics, University of Alabama,\\ Tuscaloosa,
Alabama 35487, USA\\}

\end{center}

\vskip 0.5cm

\abstract{
We present a procedure for quantizing complex projective spaces $\mathbb{CP}^{p,q}$, $q\ge 1$, as well as construct   relevant star products on these spaces. The quantization  is made unique with the demand that it preserves the full isometry algebra of the metric. Although the isometry algebra, namely $su(p+1,q)$, is preserved by the quantization, the Killing vectors generating these isometries pick up quantum corrections.   The quantization procedure is an extension of one applied recently to Euclidean $AdS_2$, where it was found that all quantum corrections to the Killing vectors vanish in the asymptotic limit, in addition to the result that the star product  trivializes  to pointwise product in the limit.  In other words, the space is asymptotically anti-de Sitter making it a possible candidate for the $AdS/CFT$ correspondence principle.   In this article, we find indications that the results for quantized Euclidean $AdS_2$ can be extended to quantized $\mathbb{CP}^{p,q}$, i.e., noncommutativity is restricted to a limited neighborhood of some origin, and these quantum spaces approach $\mathbb{CP}^{p,q}$ in the asymptotic limit.}

\section{Introduction}

{The $AdS/CFT$ correspondence principle posits {strong/weak} duality between the quantum gravity in the bulk of an asymptotically  anti-de Sitter ($AdS$) space and a conformal field theory (CFT) on the  boundary of this space. \cite{Maldacena:1997re,Aharony:1999ti} For obvious reasons, however, most practical applications of the correspondence principle utilize {\it classical} gravity in the bulk. Even though a fully consistent  quantum theory of gravity remains out of reach,} there are {model independent} indications that any theory of quantum gravity will require a quantization of spacetime {\cite{Bronstein:2012zz,Doplicher:1994tu,Doplicher:1994zv}}.  The quantization of $AdS$, or more generally  asymptotically $AdS$, spacetimes has  been examined in two dimensions,~\cite{Ho:2000fy,Fakhri:2011zz,Jurman:2013ota,Stern:2014aqa,Chaney:2015ktw} and four dimensions.\cite{Steinacker:2019fcb} Its application to  the correspondence principle has received  only some initial work in two dimensions.\cite{Pinzul:2017wch,deAlmeida:2019awj}

While in this article we do not directly address the quantization of general $AdS$ spaces of dimension larger than two, we do present a procedure for quantizing another set of non-trivial non-compact geometries generalizing the two dimensional case, namely indefinite complex projective spaces in arbitrary dimensions, $\mathbb{CP}^{p,q}$, $q\ge 1$.  We also introduce  relevant star products for these spaces. $\mathbb{CP}^{p,q}$ is a non-compact version of $\mathbb{CP}^{n}$. The simplest example of an indefinite complex projective space is $\mathbb{CP}^{0,1}$, which is equivalent to  two dimensional anti-de Sitter space, or more precisely Euclidean anti-de Sitter space, $EAdS_2$.   Another example is $\mathbb{CP}^{1,2}$, which is an $S^2$ bundle over $AdS_4$.\cite{Steinacker:2019fcb}
{While the noncommutative generalization of the compact $\mathbb{CP}^{n}$ has received some attention \cite{Balachandran:2001dd}, the same cannot be said about the non-compact case, or other non-trivial non-compact spaces. Hasebe has done a  study of quantized, or `fuzzy', hyperboloids,\cite{Hasebe:2012mz} while Steinacker and Sperling have applied such spaces, more specifically the fuzzy four-hyperboloid,  or noncommutative $AdS_4$, to quantum cosmology. The    quantization  in \cite{Steinacker:2019fcb} is made unique with the demand that it preserves the full isometry algebra of the metric of the four-hyperboloid. An isometry preserving quantization and star product can also be constructed for a general $\mathbb{CP}^{p,q}$, as we demonstrate here.  Although the isometry algebra, namely $su(p+1,q)$, is preserved by the quantization, the isometry generators, i.e., the Killing vectors, can pick up quantum corrections.}

As stated above, the simplest example of an indefinite complex projective space is $\mathbb{CP}^{0,1}$, or $EAdS_2$.   Its  isometry preserving quantization, which we denote by $ ncEAdS_2$, has been examined previously.~\cite{Ho:2000fy,Fakhri:2011zz,Jurman:2013ota,Stern:2014aqa,Chaney:2015ktw,Pinzul:2017wch,deAlmeida:2019awj} Among the results found in this case is the fact that the star product (when expressed in a suitable set of coordinates) approaches the point-wise product in the asymptotic limit (which corresponds to the boundary limit of anti-de Sitter space).~\cite{Pinzul:2017wch}  It was also argued that the quantum corrections to the Killing vectors vanish in this limit. Thus $ ncEAdS_2$ asymptotically approaches {commutative} anti-de Sitter space.  In other words, the quantum features of $ ncEAdS_2$ occur, for all practical purposes,  in a limited neighborhood of some origin. Since $ ncEAdS_2$ is an asymptotically anti-deSitter space it can then be of relevance with regard to the $AdS/CFT$ correspondence principle, which posits that for every asymptotically anti-de Sitter space there is a strong/weak duality correspondence between a bulk theory and a conformal field theory living on the conformal boundary. According to the correspondence principle, the isometries of anti-de Sitter space  are  mapped to conformal symmetries of the $CFT$ on the $AdS$ boundary. It is then reasonable to speculate that it has a conformal dual, barring  known difficulties of the correspondence principle for two dimensional anti-de Sitter space (see for example,~\cite{Strominger:1998yg,Maldacena:2016hyu}).  This was pursued in \cite{Pinzul:2017wch,deAlmeida:2019awj} where correlation functions were computed on the boundary.

As we argue in this article, the quantization procedure for $EAdS_2$ can be extended to any $\mathbb{CP}^{p,q}$, $q\ge 1$.  We can ask whether analogous conclusions can be reached regarding their asymptotic behavior. {The question therefore is whether there is a quantized version of $\mathbb{CP}^{p,q}$ which asymptotically becomes commutative. In other words: 1) Does the star product between two functions with support ``near the boundary'' reduce the commutative one, and 2) do the noncommutative corrections to the Killing vectors vanish in the boundary limit?} {Of course, ``the boundary'' refers here to the asymptotic $\mathbb{CP}^{p,q}$ region, rather than a sharp edge of the manifold.} The results obtained here do indeed support the affirmative answer to these questions. For the examples we consider  we find that, in the asymptotic limit, the relevant star product trivializes to the commutative product and noncommutative corrections to the Killing vectors vanish.

In Section 2 we review the quantization of Euclidean $AdS_2$.  We parametrize the manifold in terms of two different sets of coordinates (which differ from those used in~\cite{Pinzul:2017wch,deAlmeida:2019awj}), specifically, local affine coordinates and canonical coordinates.  The former have the advantage that they can be applied to any  complex projective space.   The  canonical coordinates, on the other hand, are useful for the purpose of quantization, and satisfy  three requirements: The first, is of course, the requirement that they obey the canonical Poisson brackets.  The second, which  is surprisingly non-trivial to  ensure,  is that they cover the entire complex plane. Dropping this condition would  necessitate a careful treatment of the boundary of the domain in the quantum theory~\cite{Pinzul:2001qh,Lizzi:2003ru}. The boundary is never a sharp one, the domain of definition is always an open set, {but when the coordinates are such that the boundary is at the finite value of these coordinates} the quantization scheme we are using cannot be applied. The third requirement  is that the  geometric measure is identical, up to a factor,  to the integration measure of standard coherent states in the resulting quantum theory.  In  this regard, the quantum theory, and corresponding coherent states, naturally follow from canonical  quantization of the canonical Poisson brackets. {We quantize the space with the introduction of a noncommutative star product of the Wick-Voros type, constructed from coherent states. We show that the product asymptotically goes to the point-wise product after re-expressing it in terms of local affine coordinates. A crucial point concerns the symmetries, implemented by the} analogues of the Killing vectors, which as stated above, preserve the full isometry algebra, here $su(1,1)$.  We perform a perturbative expansion (with respect to the quantization parameter) for the symmetry generators and compute the leading order corrections to the Killing vectors. In agreement with  results in ~\cite{Pinzul:2017wch,deAlmeida:2019awj}, these corrections are seen to vanish in the asymptotic limit. {The Wick-Voros product lends itself naturally to a matrix approximation, and considering finite matrices is tantamount to the imposition of a cutoff geometry~\cite{DAndrea:2013rix, Connes:2020ifm}, which provides both an ultraviolet and an infrared cutoff. We do not do the finite matrix approximation here.}

We review $\mathbb{CP}^{p,q}$ in  Section 3, along with its parametrization in terms of local affine coordinates and canonical coordinates.  The quantization procedure outlined above for $EAdS_2$ naturally extends to $\mathbb{CP}^{p,q}$. We do not have a universal expression for the Darboux map from local affine coordinates that is valid for all $p$ and $q$, and instead present the map for specific examples.  The examples are the two 4-(real)-dimensional indefinite complex projective spaces, $\mathbb{CP}^{1,1}$   and $\mathbb{CP}^{0,2}$, in Section 4 and 5, respectively, along with their higher dimensional analogues given in  Section 6.  Like with $EAdS_2$, the canonical coordinates obey the canonical Poisson brackets,  cover all of $\mathbb{C}^{p,q}$, and the resulting  geometric measure is proportional  to the integration measure of standard coherent states in the  quantum theory.  We carry out the quantization explicitly for the examples in Section 4 and 5, and show, like with $ncEAdS_2$, that upon taking the asymptotic limit, the star product trivializes to the commutative product and quantum corrections to the Killing vectors vanish. These quantum spaces are thus asymptotically  $\mathbb{CP}^{1,1}$   and $\mathbb{CP}^{0,2}$, respectively.  Some concluding remarks are given in Section 6.

\section{Quantization of Euclidean $AdS_2$}
\setcounter{equation}{0}

\subsection{Euclidean $AdS_2$}\label{SectionEAdS2}

To define  $AdS_2$, or its Euclidean counterpart, $EAdS_2$, it is convenient to  first introduce a three-dimensional  Minkowski background  $\mathbb{R}^{2,1}$, which we shall coordinatize  with $x_\alpha,\;\alpha=1,2,3$,  using the metric diag$(+,+,-)$.  The spaces $AdS_2$, or    $EAdS_2$, results from constraining the $SO(2,1)$ invariant $x_1^2+x_2^2-x_3^2$ to be a constant, associated with the scale.   The $AdS_2$ surface corresponds to a positive constant, while  $EAdS_2$ corresponds to a negative constant.
We shall restrict our attention in this section to the Euclidean case, as this has been of traditional interest  for the $AdS/CFT$ correspondence.  Therefore we take
\be
x_1^2+x_2^2-x_3^2=-1\;,\label{cnstfrH2}
\ee
where for convenience we fixed the scale to be one. The surface identified by this relation is a two sheeted hyperboloid. The reason why it is called Euclidean $AdS_2$ is that the induced metric has a Euclidean signature.\footnote{{For example, in the so-called global coordinates the induced metric takes the form: $$ds^2|_{EAdS} = \cosh^2 \rho dt^2 + d\rho^2 \ .$$}} Later we shall restrict to {a single component of the} hyperboloid $H^2$. This space is maximally isotropic, and the three Killing vectors, which we denote by $K_\alpha$, $\alpha=1,2,3,$ form a basis for an $so(2,1)$ algebra
\be
[K_1,K_2]=-2 K_3\ ,\qquad [K_2,K_3]=2 K_1\ ,\qquad [K_3,K_1]=2 K_2\ .\label{Kvsso21}
\ee

Because $H^2$ could be thought of as a co-adjoint orbit, a natural Lie-Poisson structure exists on it. It is easily defined by setting the Poisson brackets of the embedding coordinates {to satisfy} the $so(2,1)$ algebra
\be
\{x_1,x_2\}=-2 x_3\ ,\qquad \{x_2,x_3\}=2 x_1\ ,\qquad \{x_3,x_1\}=2 x_2\ .\label{so21pbs}
\ee
With such a choice, one can then use {Lie-}Poisson structure to implement the action  of the Killing vectors on arbitrary functions $f$ on $H^2$.
Specifically, if one defines $K_\alpha $ acting on $f$ by
\be
[K_\alpha f](x)=\{x_\alpha,f\}\label{dffKcs}\;,
\ee
then from the Jacobi identity, one  recovers $so(2,1)$ algebra of the Killing vectors~(\ref{Kvsso21}).

\subsection{Local coordinates}

A number of  coordinatizations have been introduced to $EAdS_2$. A popular choice has been Fefferman-Graham coordinates\cite{Feff} because of its convenience in the $AdS/CFT$ correspondence principle.  Here, we shall instead work with two other sets of coordinates, local affine coordinates and canonical coordinates. The former has the advantage that it can be applied to any {non-compact} projective space, while the latter provides a useful step for quantization. Although the local affine coordinates for $EAdS_2$ are not defined on the entire complex plane, {it is expedient, for the purpose of quantization, that} the canonical coordinates  span all of $\mathbb{C}$. We shall make this requirement below.  Note that the canonical coordinates we use here differ from those used in \cite{Pinzul:2017wch,deAlmeida:2019awj}, because the latter are not very useful for the higher dimensional generalizations. Both sets of coordinates are, of course, related by a canonical transformation.

\subsubsection{Local affine coordinates}

We denote the local affine coordinate of $H^2$ by $\zeta$, and its complex conjugate $\zeta^*$.  The map from the $(\zeta,\zeta^*)$ to the embedding coordinates $(x_1,x_2,x_3)$   corresponds to the non-compact analogue of a  stereographic projection of $S^2$.  It is
\be
x_1-ix_2=\frac{2\zeta}{|\zeta|^2-1}\ ,\qquad \quad  x_3=\frac{|\zeta|^2+1}{|\zeta|^2-1}\ .\label{mbdzta}
\ee
By imposing the condition   $|\zeta|>1$, we restrict to the `upper' hyperboloid, $x_3\ge 1$. $|\zeta|\rightarrow\infty$ maps the point $(x_1,x_2,x_3)=(0,0,1)$ on the  hyperboloid, while $|\zeta|\rightarrow 1$ corresponds to the asymptotic limit.
Starting with the Lorentz metric on  $\mathbb{R}^{2,1}$, and using
~(\ref{mbdzta}), we obtain the following induced metric on $H^2$
\be
ds^2\;=\frac{{4}|d \zeta|^2}{(|\zeta|^2-1)^2}\ .\label{AdS2mtrc}
\ee
This is the Fubini-Study metric, and as was indicated above, it has {Euclidean signature}.  The metric tensor $g_{\zeta,\zeta^*}=\frac{{2}}{(|\zeta|^2-1)^2}$ can be expressed in terms of the K\"ahler potential $g_{\zeta,\zeta^*}=\frac{\partial^2}{\partial\zeta\partial\zeta^*}V\,$, $\;V=-{2}\ln(|\zeta|^2-1)$. The geometric measure resulting from this  metric is
\beqa
d\mu_{\tt geom}(\zeta,\zeta^*)&=&\frac{{2}}{(|\zeta| ^2-1)^{2}}\; d\zeta \wedge d\zeta^* \ .\label{2dgammsr}
\eeqa

Using~(\ref{mbdzta}), the $so(2,1)$  Poisson brackets algebra of the embedding coordinates
(\ref{so21pbs}) results from the following fundamental Poisson bracket on $H^2$,
\be \{\zeta,\zeta^*\}=i(|\zeta|^2-1)^2 \ .\label{toosvntn}\ee
Then from~(\ref{dffKcs}) we  get  explicit expressions for the Killing vectors  in terms of the local affine coordinates
\beqa  K_1-iK_2&=&2i    \Bigl(\zeta^2 \frac\partial{\partial \zeta}-\frac\partial{\partial \zeta^*}\Bigr)\ ,
\cr&&\cr  K_3&=&2i\Bigl(\zeta\frac\partial{\partial \zeta}-\zeta^*\frac\partial{\partial \zeta^*}\Bigr)\ .\label{cxpfrads2Kvs}
\eeqa

\subsubsection{Canonical coordinates}

We next apply a  Darboux transformation from the local affine  coordinates to canonical coordinates $(y,y^*)$, satisfying
\be  \{y,y^*\}=-i\label{2dcnclpb}\ee
As stated above, for the purpose  of quantization it is necessary to have $y$ span all of the complex plane, unlike $\zeta$ which is defined only outside the unit disc, $ |\zeta|>1 $. This fixes $(y,y^*)$ up to canonical transformations. {For the natural ansatz $y = f(|\zeta|)\zeta$, one obtains the following condition on the function $f(x)$:
\be
f^2 + \frac{x}{2}(f^2)' = -\frac{1}{(x^2 - 1)^2}\ ,
\ee
which has the general solution
\be
f(x)^2 = \frac{C}{x^2} + \frac{1}{x^2 (x^2 -1)}\ ,
\ee
where $C$ is an arbitrary non-negative constant. From here it follows that $|y|^2=C + \frac 1{{|\zeta|^{2}-1}}$, and it spans the entire positive real axis (including $|y|=0$) only when $C=0$. Then for this ansatz, we have
\be y=\frac{\zeta}{|\zeta|\sqrt{|\zeta|^{2}-1}}\ .\label{2dDbxtrn}\ee
}

Another desirable feature, from the point of view of quantization, is that the  geometric measure reduces to a flat measure when expressed in terms of the canonical coordinates. This easily follows from the Jacobian of the transformation, which is $\Big|\frac{\partial (\zeta,\zeta^*)}{\partial (y,y^*)}\Big|{\equiv |\{\zeta ,\zeta^*\}|}=(|\zeta|^2-1)^2$. So~(\ref{2dgammsr}) is transformed to
\beqa
d\mu_{\tt geom}(y,y^*)&=&{2}\; dy \wedge dy^* \ .\label{2dfltgmmsr}
\eeqa

When re-expressed in terms of  $(y,y^*)$, the expression~(\ref{mbdzta}) for the embedding coordinates becomes
\be
x_1-ix_2=2y\sqrt {|y|^2+1}\ ,\qquad \quad  x_3=2|y|^2+1\ .\label{mbdntsfys}
\ee
Therefore the origin of the complex plane spanned by the canonical coordinates {is the image of} the point $(x_1,x_2,x_3)=(0,0,1)$ on the  hyperboloid, while $|y|\rightarrow\infty$ corresponds to the asymptotic limit. The Killing vectors~(\ref{cxpfrads2Kvs}) when expressed in terms of the canonical coordinates become
\beqa\label{cxpfrads2Kvscc}
K_1-iK_2&=&\frac{i}{\sqrt{|y|^2+1}}\,\Bigl(y^2\frac\partial{\partial y}-(2+3|y|^2)\frac\partial{\partial y^*}\Bigr)\ ,
\cr&&\cr  K_3&=&2i\Bigl(y\frac\partial{\partial y}-y^*\frac\partial{\partial y^*}\Bigr)\ .
\eeqa

\subsection{Quantization}

One can now perform canonical quantization by replacing the   coordinates $(y,y^*)$ by   operators   $(\hat y,\hat y^\dagger)$ satisfying commutation relations
\be
[\hat y,\hat y^\dagger]=\kbar\,\BI\;,\label{2dcrsfrzzdgr}
\ee
$\kbar$ being the noncommutative parameter, and $\BI$ the identity operator.  Equivalently, we have raising and lowering operators, $\hat a^\dagger={\hat y^\dagger}/ {\sqrt{\kbar}}$ and $\hat a={\hat y}/ {\sqrt{\kbar}}$, satisfying $[\hat a,\hat a^\dagger]=\BI$. {Note that, apart from the commutation relation, it is equally fundamental  that the canonical coordinates  $y,y^*$ were defined on the whole plane} {(unlike the case with $\zeta,\zeta^*$). Otherwise, one would require a delicate treatment of the domain with a boundary.\cite{Pinzul:2001qh,
Lizzi:2003ru}

The operators $\hat y$ and $\hat y^\dagger$ act on the infinite-dimensional harmonic oscillator Hilbert space ${\cal H}$  spanned by orthonormal  states $| n\rangle,\;n=0,1,2...$
\be {| n\rangle}=\frac{(\hat a^\dagger)^{n}}{\sqrt{n!}}| 0\rangle\;,\label{2dstsndsctrp}
\ee
where $\hat a|0\rangle=0 $,  and $\langle 0| 0\rangle=1$.  Alternatively, one can introduce standard coherent states $\{| \alpha\rangle\in {\cal H},\;\alpha\in\mathbb{C}\}$   written on  $\mathbb{C}$:
\be | \alpha\rangle =e^{-\frac {|\alpha|^2}2}e^{\alpha \hat a^\dagger}| 0\rangle\;,\label{2dchrntstt}\ee
where  $\alpha $ is the eigenvalue  of $\hat a$, $\hat a |\alpha\rangle=\alpha|\alpha\rangle$.
Coherent states  form an over-complete set with unit norm.
The completeness relation and normalization condition are
\beqa
&&\int d\mu(\alpha,\alpha^{\,*})| \alpha\rangle\langle \alpha |=\BI\ ,\cr &&\cr &&
\langle \alpha| \alpha'\rangle=\exp\Bigl\{{\alpha^*{\alpha'}-\frac {|\alpha|^2}2-\frac {|\alpha'|^2}2\Bigr\}}\ .\label{2dcompnorm}
\eeqa
The  integration measure for coherent states  $d\mu(\alpha,{\alpha}^{\,*})$ is
\be
d\mu(\alpha,{\alpha}^{\,*})=\frac i{2\pi}\,d\alpha\wedge d\alpha^*=\frac i{2\pi\kbar}\,dy\wedge d y^*\;,\label{2dmsrfrchsts}
\ee
which is, up to a factor,  identical to the geometric measure~(\ref{2dfltgmmsr}). Here we have re-introduced the canonical coordinates  $(y,y^*)$ using  ${ y} ={\sqrt{\kbar}}\alpha$ and ${ y^*} ={\sqrt{\kbar}}\alpha^*$.

The {Wick}-Voros star product, $\star$, is constructed from the standard coherent states. Here we briefly review it. For details of the construction see, e.g.\ \cite{Alexanian:2000uz,Zachos:2000zh,Galluccio:2008wk}. One first defines  symbols  ${\cal A}(\alpha,\alpha^{*})$  on  the complex plane associated with operator functions  $A$ of $\hat a$ and $\hat a^\dagger$ using
\be
{\cal A} (\alpha,\alpha^{*})=\langle \alpha| A|\alpha\rangle\ .\label{2dvrssmbl}
\ee
Then given any two functions $A$ and $B$  of $\hat a$ and $\hat a^\dagger$, with symbols ${\cal A}$ and ${\cal B}$, respectively,
the  symbol  of their product is
\be
[{\cal A}\star{\cal B}](\alpha,\alpha^{\,*})=\langle \alpha|AB| \alpha\rangle\label{smblfprdct}\ ,
\ee
which gives  the {Wick}-Voros star product of the two symbols. It is given explicitly in terms of the canonical coordinates by
\be
[ {\cal A}\star {\cal B}](y,y^*)={\cal A}(y,y^*)\,\exp \Bigl\{\kbar \overleftarrow{\frac\partial{\partial y}}\;\overrightarrow{\frac{ \partial}{\partial y^*}}\Bigr\}\,{\cal B}(y,y^*)\ .\label{2dVrsstprd}
\ee
This expression realizes the fundamental commutation relation,  $ [ y, y^*]_\star=\kbar$, where $[ {\cal A},{\cal B}]_\star={\cal A}\star {\cal B}-{\cal B}\star {\cal A}$ denotes the star commutator, and gives the desired commutative limit,
\beqa
{\cal A}\star {\cal B}&=&{\cal A} {\cal B}+ {\cal O}(\kbar)\ ,\cr
[ {\cal A},{\cal B}]_\star&=&i\kbar \{ {\cal A},{\cal B} \}+ {\cal O}(\kbar^2)\ .\label{clmtfstrctr}
\eeqa

The star product can be re-expressed in terms of the local affine coordinates using~(\ref{2dDbxtrn}). One gets
\be
[ {\cal A}\star {\cal B}](\zeta,\zeta^*) = {\cal A}(\zeta,\zeta^*)\,\exp\Bigl\{\kbar  \overleftarrow{{\cal D}}  \overrightarrow{ {\cal D}^*}\Bigr\}\,{\cal B}(\zeta,\zeta^*)\;,
\ee
where
\beqa
{\cal D}&=&\frac{\sqrt{{|\zeta|^2-1}}}{2|\zeta |}\Bigl( {\frac\partial{\partial \zeta}}-(2|\zeta|^2-1)\frac{\zeta^*}{\zeta}\, {\frac\partial{\partial \zeta^*}}\Bigr)\ ,\cr
&&\cr {\cal D}^*&=&\frac{\sqrt{{|\zeta|^2-1}}}{2|\zeta |}\Bigl( {\frac\partial{\partial \zeta^*}}-(2|\zeta|^2-1)\frac{\zeta}{\zeta^*}\, {\frac\partial{\partial \zeta}}\Bigr)\ .
\eeqa
{The presence of the $\sqrt{|\zeta|^2-1}$ factor is crucial. As we mentioned earlier, the {conformal} boundary is obtained in the limit $|\zeta|\rightarrow 1$, and therefore this shows that the value of the product of two functions asymptotically is not different from the one obtained with the usual commutative multiplication.} {(Provided the noncommutative corrections to the functions vanish at the conformal boundary.)} It also means that the star commutator reduces to $i\kbar$ times the Poisson bracket  in the asymptotic limit.

We will characterize noncommutative $EAdS_2$ in terms of the  noncommutative analogues of the embedding coordinates $(x_1,x_2,x_2)$~\cite{Ho:2000fy,Fakhri:2011zz,Jurman:2013ota,Stern:2014aqa,Chaney:2015ktw}. We need a set of noncommutative}coordinates, which we call $X_\alpha$, that satisfy the $\star$ analogue of the conditions~(\ref{cnstfrH2}) and~(\ref{so21pbs}):
\be
X_1\star X_1+X_2\star X_2-X_3\star X_3=-{\cal C}\label{capXcnst}
\ee
and
\be
[X_1,X_2]_\star=-2i\kbar X_3\ ,\qquad   [X_2,X_3]_\star=2 i\kbar X_1\ ,\qquad  [X_3,X_1]_\star=2i\kbar X_2\label{cmtrmbdops}\ ,
\ee
with ${\cal C}>0$, a constant which defines the Euclidean version of noncommutative $AdS_2$. In order to recover~(\ref{cnstfrH2}) in the commutative limit, we need  ${\cal C}=1+{\cal O}(\kbar)$. The $X$'s should be functions of the embedding coordinates $(x_1,x_2,x_3)$ of the commutative theory, and must reduce to them in the limit {(or, in terms of the local coordinates, they should be functions of $(\zeta, \zeta^*)$ or $(y,y^*)$ and must reduce to (\ref{mbdzta}) or (\ref{mbdntsfys}) respectively)}. Relation~(\ref{cmtrmbdops}) for the $X_\alpha$'s then defines the $so(2,1)$  algebra, and  ${\cal C}$ fixes the Casimir.  We thereby obtain irreducible representations of $so(2,1)$.

Given  the  noncommutative analogues of the embedding coordinates, one can introduce noncommutative analogues of the Killing vectors of $EAdS_2$. Denote them by $K^\star_\alpha$.  They are defined in analogous way to $K_\alpha$, by essentially replacing the Poisson bracket in~(\ref{dffKcs}) by the star commutator:
\be
[K^\star_\alpha f](X)=\frac 1{i\kbar}[X_\alpha,f]_\star\label{ncdffKcs}\ ,
\ee
where  $f(X)$ denotes a function on $ncEAdS_2$.  Like the Killing vectors $K_\alpha$ of $EAdS_2$,  $K^\star_\alpha$  satisfy the  $so(2,1)$ algebra. Furthermore, from~(\ref{clmtfstrctr}), we see that $K^\star_\alpha$ reduce to $K_\alpha$ in the commutative limit.  On the other hand, the expressions~(\ref{cxpfrads2Kvs}) for $K_\alpha$ do not hold for the noncommutative analogues of the Killing vectors (except for $\alpha =3$, and except for the asymptotic limit, as we shall see below). Thus, quantization leads to  deformations of the Killing vectors, although the algebra they generate is not deformed.

{We next write  $X_\alpha$ in terms of the canonical coordinates $y$ and $y^*$. For this we will need several simple properties of the star product (\ref{2dVrsstprd}).}

\begin{enumerate}
\item The symbol of the operator $\hat{y}^\dagger \hat{y}$ is $|y|^2$. In general, any function $\mathcal{F}(|y|^2)$ is a symbol of some operator $F(\hat{y}^\dagger \hat{y})$ and vice versa, any operator $F(\hat{y}^\dagger \hat{y})$ has a symbol depending only on $|y|^2$:
\be\label{symbolF}
\mathcal{F}(|y|^2) = \exp \left(-\frac{|y|^2}{\kbar}\right)\sum\limits_{n=0}^{\infty} \frac{|y|^{2n}}{\kbar^n n!}F(\kbar n) \ .
\ee

\item For any function $\mathcal{F}(y,y^*)$, we have
\be\label{FyyF}
\mathcal{F}(y,y^*)\star y = y\mathcal{F}(y,y^*)\ ,\qquad y^* \star \mathcal{F}(y,y^*) = y^* \mathcal{F}(y,y^*)
\ee
(The ordering on the left hand side of the equations is important.)

\item For any two functions of $|y|^2$, $\mathcal{F}(|y|^2)$ and $\mathcal{G}(|y|^2)$, we have
\be\label{FstarG}
\mathcal{F}(|y|^2)\star \mathcal{G}(|y|^2) = \sum\limits_{n=0}^{\infty} \frac{\kbar^n |y|^{2n}}{n!}\mathcal{F}^{(n)}(|y|^2) \mathcal{G}^{(n)}(|y|^2)\ ,
\ee
where the derivative is taken with respect to $|y|^2$.
\end{enumerate}

Motivated by (\ref{mbdntsfys}), we look for the noncommutative coordinates $X_\alpha$ satisfying (\ref{cmtrmbdops}) in the form
\be\label{2dnstz}
X_{-}=X_1-iX_2=2{\cal S}\star y\stackrel{(\ref{FyyF})}{\equiv} 2y{\cal S}\ ,\qquad   X_{+}=X_1+iX_2=2 y^* \star{\cal S}\stackrel{(\ref{FyyF})}{\equiv} 2y{\cal S} \ ,
\ee
where ${\cal S}={\cal S}(|y|^2)$ is some  real function to be determined below. Using the properties of the star product (\ref{symbolF}-\ref{FstarG}), one can easily find
\be\label{XplusXminus}
[X_{-},X_{+}]_\star = 4\kbar \left( \mathcal{S}\star \mathcal{S} + |y|^2 (\mathcal{S}\star \mathcal{S})'\right)\ ,
\ee
where the prime denotes a derivative with respect to $|y|^2$. According to (\ref{cmtrmbdops}) this should be equal to $4\kbar X_3$. So we have that $X_3 = X_3 (|y|^2)$, and in terms of $\mathcal{S}$ is given by
\be\label{X3}
X_3 = \mathcal{S}\star \mathcal{S} + |y|^2 (\mathcal{S}\star \mathcal{S})' \ .
\ee
{Using (\ref{cmtrmbdops}) one more time
\be\label{XminusX3}
4\kbar \mathcal{S}\star y \equiv 2\kbar X_{-} = [X_{-} , X_3]_\star = 2\mathcal{S}\star [y , X_3]_\star
\ee
and taking into account that there exists $\mathcal{S}^{-1}$ such that $\mathcal{S}^{-1}\star\mathcal{S} = 1$ (since it exists to zeroth order in $\kbar$, and we assume that the expansion in $\kbar $ is valid) we arrive at the equation for $X_3$
\be\label{X3equation}
[y , X_3]_\star = 2\kbar y\ \ \mathrm{or}\ \ {X_3}' = 2\ ,
\ee
which leads to
\be\label{xtreelnr}
X_3 =2|y|^2+c\ ,\ c={\rm constant}\ .
\ee
Using this in (\ref{X3}) we arrive at the differential equation for $\mathcal{S}\star\mathcal{S}$
\be\label{S2equation}
\mathcal{S}\star \mathcal{S} + |y|^2 (\mathcal{S}\star \mathcal{S})' = 2 |y|^2 +c\ ,
\ee
which is easily solved to give
\be
\mathcal{S}\star\mathcal{S} = |y|^2 +c + \frac{a}{|y|^2}\ ,
\ee
where $a$ is another integration constant. It is clear that one should set $a=0$ in order to have non-singular noncommutative corrections for $|y|\rightarrow 0$ (and to recover that $X_1,\,X_2 \rightarrow 0$ in this limit). So, we have
\be\label{S2}
\mathcal{S}\star\mathcal{S} = |y|^2 +c\ .
\ee}
{The Casimir in (\ref{capXcnst}) is now easily computable
\be\label{Casimir}
\mathcal{C} =- \frac{1}{2}\left( X_{-}\star X_{+} + X_{+}\star X_{-} \right) + X_{3}\star X_{3} = c^2 - 2\kbar c \ .
\ee}

{In general, the constant $c$ should have the form $c = 1 + \mathcal{O}(\kbar)$. We fix this freedom in quantization by requiring that the symbol $X_3$ remains undeformed, i.e. by setting $c=1$. Then (\ref{S2}) looks exactly as in the commutative case (\ref{mbdntsfys})
\be\label{S2c1}
\mathcal{S}\star\mathcal{S} = |y|^2 +1\ ,
\ee
i.e. $\mathcal{S}$ is a symbol of the operator $\sqrt{1 + \hat{y}^\dagger \hat{y}}$, which can be formally written using (\ref{symbolF}) as
\be\label{Sseries}
\mathcal{S}(|y|^2) = \exp\left( -\frac{|y|^2}{\kbar} \right)\sum\limits_{n=0}^{\infty}\frac{|y|^{2n}}{\kbar^n n!}\sqrt{\kbar n +1}\ .
\ee}

{Though we do not have the closed answer for the series (\ref{Sseries}), we can systematically calculate $\mathcal{S}$ to any order in $\kbar$. Let $\mathcal{S}_n$ be the functions independent of $\kbar$ and defined by
\be\label{Sn}
\mathcal{S}(|y|^2) = \sum\limits_{n=0}^{\infty} \kbar^n \mathcal{S}_n (|y|^2)\ .
\ee
Plugging this into (\ref{S2c1}) and using (\ref{FstarG}) we have after some trivial index relabeling
\be\label{Snseries}
1 + |y|^2 = \mathcal{S}\star\mathcal{S} = \sum\limits_{n=0}^{\infty} \kbar^n \left( \sum\limits_{m=0}^{n}\frac{|y|^{2m}}{m!} \sum\limits_{r=0}^{n-m} \mathcal{S}^{(m)}_{n-m-r} \mathcal{S}^{(m)}_{r} \right)\ .
\ee
From (\ref{Snseries}) we obtain the recursion relations defining $\mathcal{S}_n$ for any $n$:
\beqa\label{Snrecursion}
n=0\ &,&\ \mathcal{S}_0 = \sqrt{1 + |y|^2} \nonumber \\
n\geq 1\ &,& \ \sum\limits_{m=0}^{n}\frac{|y|^{2m}}{m!} \sum\limits_{r=0}^{n-m} \mathcal{S}^{(m)}_{n-m-r} \mathcal{S}^{(m)}_{r} = 0 \ .
\eeqa
For example, for $n=1$ we have
\be\label{n1}
\mathcal{S}_1 \mathcal{S}_0 + \mathcal{S}_0 \mathcal{S}_1 + |y|^2 {\mathcal{S}_0}'{\mathcal{S}_0}' = 0\ \Rightarrow\ \mathcal{S}_1 = -\frac{|y|^2}{8(1 + |y|^2)^{3/2}}\ .
\ee
In general, it is not hard to see from (\ref{Snrecursion}) that for an arbitrary $n$, $\mathcal{S}_n $ will have the following form
\be\label{Snstructure}
\mathcal{S}_n = \sqrt{1 + |y|^2}\frac{P_{n}(|y|^2)}{(1 + |y|^2)^{2n}} =: \sqrt{1 + |y|^2} \mathcal{L}_n\ ,
\ee
where $P_{n}(x)$ is some polynomial of degree $n$, with $P_0=1$. Then we can write our noncommutative coordinates $X_\alpha$ in terms of the commutative ones as
\be\label{Xx}
X_{\pm} = x_{\pm} \sum\limits_{n=0}^{\infty} \kbar^n \mathcal{L}_n\ ,\ X_3 = x_3 \ ,
\ee
 $x_\pm=x_1\pm x_2$ being  the commutative counterparts  to $X_\pm$.  We conclude that $X_\pm\rightarrow x_\pm$ in the asymptotic limit $|y|^2 \rightarrow \infty$,
\be\label{Xxasympt}
X_{\pm} = x_{\pm} \left( 1 + \mathcal{O}\Bigl(\frac{\kbar}{|y|^2}\Bigr) \right)\ ,\ X_3 = x_3\ .
\ee}

{Using (\ref{Xx}) and its asymptotics (\ref{Xxasympt}) we can easily study the behaviour of the noncommutative Killing vectors, defined by (\ref{ncdffKcs}), near the conformal boundary. Let us denote by $\mathcal{L}$ the sum in (\ref{Xx}), $ \mathcal{L}=\sum\limits_{n=0}^{\infty} \kbar^n \mathcal{L}_n$.  Since $X_3=x_3$,  $K^\star_3$ has exactly the same form as its commutative counterpart $K_3$ in (\ref{cxpfrads2Kvscc}). Trivial analysis shows that when $|y|\rightarrow\infty$, $K^\star_{\pm}$ behave as
\beqa\label{Kasympt}
&& K^\star_{\pm}f(y,y^*) \equiv  \frac{1}{i\kbar} [X_{\pm}, f]_\star = \nonumber\\
&& =\mathcal{L}K_{\pm} f + \frac{1}{2}x_\pm\mathcal{L}' K_3 f + \sum\limits_{n=2}^{\infty}\frac{(i\kbar)^{n-1}}{n!} \left[ \partial^{n}_y (x_{\pm}\mathcal{L}) \partial^{n}_{y^*} f - \partial^{n}_{y^*} (x_{\pm}\mathcal{L}) \partial^{n}_{y} f \right] =  \nonumber\\
&& = \left( 1 + \mathcal{O}\Bigl(\frac{\kbar}{|y|^2}\Bigr) \right) K_{\pm} f \ ,
\eeqa
where we naturally assumed that $K_3 f$ has the same asymptotic behavior as $K_{\pm} f$. This shows that the noncommutative corrections to  $ K^\star_\alpha$ vanish in the asymptotic limit. Of course, the same is true for the case of the local affine coordinates $(\zeta, \zeta^*)$. In this case the commutative limit for both, the coordinates $X_\alpha$ and Killings $K^\star_\alpha$, will be recovered as $|\zeta|^2\rightarrow 1$.}

{Thus upon expressing the system in terms of the canonical or local affine coordinates, we see that the noncommutative coordinates $X_\alpha$ as well as the $so(2,1)$ isometry generators of $ncEAdS_2$ approach the standard $EAdS_2$ expressions, while the  star product approaches the ordinary product, which is seen in local affine coordinates. We can then argue that $ncEAdS_2$ reduces to $EAdS_2$ in the asymptotic limit.}

\section{$\mathbb{CP}^{p,q}$}
\setcounter{equation}{0}

The natural generalization of  $ncEAdS_2$ is the quantization of  the indefinite complex projective space, denoted by $\mathbb{CP}^{p,q}$, where $p$ and $q$ are positive integers; $p$ can be zero, while $q\ge 1$.   $EAdS_2$  corresponds to $p=0$, $q=1$.  In this section we review $\mathbb{CP}^{p,q}$, writing down the Killing vectors and  analogues of embedding coordinates in terms of appropriate Fubini-Study coordinates $( \zeta^i,\zeta^*_i),\; i=1,...,p+q$,  for these spaces. In order to reproduce  the quantization program of the previous section, we will need to find the Darboux transform from the Fubini-Study coordinates to canonical coordinates $(y_i,y_i^*)$ spanning {\it all} of $ \mathbb{C}^{p+q}$.  As was mentioned for the case of $EAdS_2$, if the canonical coordinates do not span the entire $ \mathbb{C}^{p+q}$, quantization becomes {unmanageable} due to the presence of boundaries.  We have not found a general expression for the Darboux transformation that applies to all  $\mathbb{CP}^{p,q}$ spaces.  Rather, we can give the transformation for various classes of such spaces, which we shall illustrate in Sections 4 and 5.

\subsection{Definition}

The space  $\mathbb{CP}^{p,q}$, $q\ge 1$,  is defined as the $\mathbb{H}^{2q,2p+1}$ hyperboloid mod $S^1$.  It can be constructed  starting from  a $p+q+1$ dimensional complex  space $\mathbb{C}^{p+1,q}$, with indefinite metric
\be
\eta_{\mathbb{C}}={\rm diag}(\underbrace{+...+}_{p+1},\underbrace{-...-}_q)\ .\label{thspnrmtrc}
\ee
Say  $\mathbb{C}^{p+1,q}$ is coordinatized by $z^a$, $a=1,...,p+q+1$, along with their complex conjugates  ${z^a}^*$, where the indices $a,b,...$ are raised and lowered using the metric  $\eta_{\mathbb{C}}$. To embed $\mathbb{H}^{2q,2p+1}$  in  $\mathbb{C}^{p+1,q}$ one   imposes  the constraint
\be
z_a^*z^a=1\ .\label{ess7}
\ee
To obtain $\mathbb{CP}^{p,q}$ one further makes the identification
\be
z^a\sim e^{i\chi} z^a\ ,\label{eqvrltn}
\ee
$ e^{i\chi}$ being an arbitrary phase.  The compact complex projective space  $\mathbb{CP}^{p}$ corresponds to $q=0$. We will not be concerned with it in the following. The space $\mathbb{CP}^{p,q}$ can be equivalently defined as the coset space $SU(p+1,q)/U(p,q)$.

The standard metric and Poisson bracket on complex projective spaces are  the  Fubini-Study metric and the canonical one, respectively.  The former  is given by
\be
ds^2=d z_a^* dz^a-| z^*_a dz^a|^2\ ,\label{FSmtrc}
\ee
while the latter is
\be
\{z^a,z^*_b\}=-i\delta^a_b\;,\qquad \{z^a,z^b\}=\{z^*_a,z^*_b\}=0\;,\qquad a,b=1,...,p+q+1\label{pbsfzzs}\ .
\ee
Using~(\ref{pbsfzzs}), it follows that~(\ref{ess7}) is the first class constraint (in the sense of Dirac's Hamiltonian formalism) that  generates the phase equivalence~(\ref{eqvrltn}).

\subsection{Coordinates}

Here we are interested in generalizing the two sets of coordinates given previously for $EAdS_2$, i.e., local affine coordinates and canonical coordinates. While here  we give explicit expressions for the former, we just discuss qualitative features of the latter. We shall postpone giving explicit expressions for the Darboux transformation to sections which follow.

\subsubsection{Local affine coordinates}
The local affine coordinates $( \zeta^i,\zeta^*_i),\; i=1,...,p+q$,  are  defined in terms of  the coordinates $z^a$ by\footnote{As usual, one can replace $ z^{p+q+1} $ in the denominator by another complex coordinate, say $ z^{\tt a} $, which would be valid for $z^{a}\ne 0$, thereby defining a  local affine coordinates on a different coordinate patch.}
\be
\zeta^i=\frac{z^i}{z^{p+q+1}}\;,\qquad z^{p+q+1}\ne 0 \ .\label{ohptwon}
\ee
 They are invariant under the phase equivalence transformation~(\ref{eqvrltn}). The $\zeta^*_i$ are obtained by taking the complex conjugate of~(\ref{ohptwon}) and lowering the index using the background metric tensor on the  $p+q\; \;$dimensional subspace~(\ref{thspnrmtrc}). We note that it is the Euclidean metric for the special case of $q=1$. From the constraint~(\ref{ess7}), one has
\be
\zeta^i\zeta^*_i=1+\frac 1{|z^{p+q+1}|^2}\ ,\label{wonpt8}
\ee
and it follows that $\zeta^i\zeta^*_i>1$, which further implies that   $\; |\zeta_1|^2+\cdots + |\zeta_{p+1}|^2>1$.    Therefore the coordinate patch spanned by  $( \zeta^i,\zeta^*_i)$ is  ${\mathbb{C}}^{p+1,q-1}$ with the region  $\zeta^i\zeta^*_i\le 1$ removed.  For reasons stated below we call the boundary of this region the general {\it asymptotic limit}:
\be
\zeta^i\zeta^*_i\rightarrow 1 \qquad {\rm or}\qquad   z^{p+q+1}\rightarrow 0\label{thesmptlmt}\ .
\ee
This is in agreement with the asymptotic limit defined previously for $EAdS_2$.

While~(\ref{thspnrmtrc}) is the background metric, the metric on the surface $\mathbb{CP}^{p,q}$ is the Fubini-Study metric~(\ref{FSmtrc}). Substituting ${z^i}={z^{p+q+1}}\,\zeta^i$ into~(\ref{FSmtrc}) gives the Fubini-Study metric tensor in terms of local affine coordinates
\be
ds^2\;=\;g_{i\bar j}(\zeta,\zeta^*)\, d\zeta^id\zeta^*_j\;=\;\frac{d \zeta_i^* d\zeta^i}{{\cal Z}^2}\;-\;\frac{|\zeta_i^*d\zeta^i|^2}{{\cal Z}^4}\;,\qquad i,j,k,...=1,...,p+q\label{FSnafncrds}\ ,
\ee
where we denote
\be
{\cal Z}^2= \zeta^i\zeta^*_i- 1\ .
\ee
For $p=0, q=1$, $g_{i\bar j}(\zeta,\zeta^*)$ reduces to the metric tensor~(\ref{AdS2mtrc}) on $EAdS_2$ {(up to an overall factor)}. It can be expressed in terms of the K\"ahler potential
\be
g_{i\bar j}=\frac{\partial^2}{\partial\zeta^i\partial\zeta^*_j}2\ln {\cal Z}\ .
\ee
The geometric measure associated with the metric~(\ref{FSnafncrds}) is
\beqa
d\mu_{\tt geom}(\zeta,\zeta^*)=\frac{1 }{2^{p+q}{\cal Z}^{2(p+q+1)}}\; d\zeta^1 \wedge \cdots\wedge d\zeta^{p+q} \wedge d\zeta^*_1\wedge\cdots \wedge d\zeta^*_{p+q}\ , \label{mugeom}
\eeqa
which is the generalization of~(\ref{2dgammsr}). To verify~(\ref{mugeom}) we only need the identity
\be
\det (\BI_n + vw^T)=1+w^Tv\ ,\label{Ayewon}
\ee
where $v,w\in {\rm Vec}_n$ and $\BI_n$ is the n-dimensional identity matrix, which easily follows from the definition of the determinant, det$M=\frac 1{n!}\epsilon_{i_1\cdots i_n}\epsilon_{j_1\cdots j_n} M_{i_1j_1}\cdots M_{i_nj_n}$ for any $M\in {\rm Mat}_n$.
We can write the invariant interval in~(\ref{FSnafncrds}) as
$$ ds^2=d\Xi^T G\, d\Xi \;, $$
\be
  G=\frac{\gamma^2}2\pmatrix{0& \BI_{p+q}-\gamma^2\zeta^*\zeta^T\cr\BI_{p+q}-\gamma^2\zeta\zeta^{*T}&0}\;,\qquad  \Xi=\pmatrix{\zeta\cr\zeta^*}\;,
\ee
where  $\zeta=\pmatrix{\zeta^1\cr : \cr\zeta^{p+q}}$,  $\zeta^*=\pmatrix{\zeta^*_1\cr : \cr\zeta^*_{p+q}}$ and $\gamma=\frac 1{{\cal Z}}$.
The geometric measure  is then
\beqa
d\mu_{\tt geom}(\zeta,\zeta^*)&=&\sqrt{| {\rm det}\,G|}\; d\zeta^1 \wedge \cdots\wedge d\zeta^{p+q} \wedge d\zeta^*_1\wedge\cdots \wedge d\zeta^*_{p+q}\ .
\eeqa
In order to  recover~(\ref{mugeom}), we then use~(\ref{Ayewon}) to get,
\be
{\rm det}\,G=-\frac{\gamma^{4(p+q)}}{2^{2(p+q)}}\Bigl( {\rm det}\,( \BI_{p+q}-\gamma^2\zeta^*\zeta^T)\Bigr)^2=-\frac{\gamma^{4(p+q+1)}}{2^{2(p+q)}}\ .
\ee
From~(\ref{pbsfzzs}), the Poisson brackets on the coordinate patch spanned by $( \zeta^i,\zeta^*_i)$ are
\be
\{\zeta^i,\zeta^*_j\}=i{\cal Z}^2(\zeta^i\zeta^*_j-\delta^i_j)\;,\ \ \{\zeta^i,\zeta^j\}=\{\zeta^*_i,\zeta^*_j\}=0 \;,\label{pbsfrztab}
\ee
generalizing the Poisson bracket~(\ref{toosvntn}) for the case of $EAdS_2$.

The  isometry group of $\mathbb{CP}^{p,q}$ is $SU(p+1,q)$. There are then a total of $(p+q)(p+q+2)$ Killing vectors associated with the metric tensor~(\ref{FSnafncrds}).  In terms of the local affine coordinates they are given by
\beqa
\kappa_i^{\;\,j}&=&\zeta^j\frac\partial{\partial \zeta^i}-\zeta^*_i\frac\partial{\partial \zeta^*_j}\ ,\qquad\qquad\qquad \cr&&\cr
\kappa_i^{\;\,p+q+1}&=&\frac\partial{\partial \zeta^i}-\zeta^*_i\zeta^*_j\frac\partial{\partial \zeta^*_j}\ ,\cr&&\cr
\kappa_{p+q+1}^{\qquad\,i}&=&\frac\partial{\partial \zeta^*_i}-\zeta^i\zeta^j\frac\partial{\partial \zeta^j}\ ,\label{KvsncmtvCP11alc}\eeqa
generalizing~(\ref{cxpfrads2Kvs}).
$\kappa_{i\,j}$, $\kappa_{i\;p+q+1}$ and  $\kappa_{p+q+1\;i}$ form a basis for $su(p+1,q)$
\beqa
[ \kappa_i^{\;\,j}, \kappa_k^{\;\,\ell}]&=&\delta _i^\ell\, \kappa_k^{\;\,j}- \delta _k^j\, \kappa_i^{\;\,\ell}\ ,\cr&&\cr
[\kappa_i^{\;\,p+q+1} , \kappa_j^{\;\,k}]&=&\delta_i^k\,\kappa_j^{\;\,p+q+1}\ ,\cr&&\cr
[\kappa_i^{\;\,j} ,\kappa_{p+q+1}^{\qquad\,k}]&=&\delta_i^k\,\kappa_{p+q+1}^{\qquad\,j}\ ,\cr&&\cr
[\kappa_i^{\;\,p+q+1} ,\kappa_{p+q+1}^{\qquad\,j}]&=&- \kappa_i^{\;\,j}-\delta^j_i\, \kappa_k^{\;\,k}\ .
\eeqa
To recover the Killing vectors $K_1,K_2,K_3$ defined previously for $EAdS_2$ we need $ K_1-iK_2=-2i \,\kappa_1^{\;\,2}$ and $K_3=2i\,\kappa_1^{\;\,1}$.

By generalizing the notion of the real embedding coordinates $x_i$ for $EAdS_2$~(\ref{mbdzta}), we can implement the action of the Killing vectors~(\ref{KvsncmtvCP11alc}) using the Poisson bracket (\ref{pbsfrztab}). Call $x_a^{\;\,b},\; a,b=1,...,p+q+1$, real embedding coordinates for $\mathbb{CP}^{p,q}$ (in contrast to the complex embedding coordinates $z^a$). Their Poisson bracket algebra  should correspond to $su(p+1,q)$ .  For this we define $x_a^{\;\,b}$ in terms of $z^a$'s and then on the coordinate patch spanned by the local affine coordinates $( \zeta^i,\zeta^*_i)$. In terms of the complex embedding coordinates we have:
\be\label{xab}
x_a^{\;\,b} = z^*_a z^b\ , \qquad a,b=1,...,p+q+1 \ .
\ee
Using (\ref{pbsfzzs}) one can easily see that the Poisson brackets of $x_{ab}$ close to give the $su(p+1,q)$ isometry algebra
\be
\{ x_{ab},  x_{cd}\}=i({\eta_{\mathbb{C}}}_{ad}\, x_{cb}- {\eta_{\mathbb{C}}}_{cb}\, x_{ad})\ .\label{psfrtrclssx}
\ee
Then, as usual, we can write the action of $SU(p+1,q)$ Killing vectors in terms of these Poisson brackets
\be
{ \kappa}_a^{\;\,b} f=-i\{x_a^{\;\,b},f\}\ .\label{threftwnx}
\ee
The appearance of an extra Killing vector due to $x_{p+q+1}^{\;\;\;\,p+q+1}$ is apparent, which could be seen by noticing that not all $x_a^{\;\,b}$'s are independent due to the constraint (\ref{ess7}), which leads to
$$ {\rm tr}\, x \;\;=\;\;x_a^{\;\,a}=1\ ,\;\qquad\qquad\quad\,$$
as well as the higher order conditions
\beqa {\rm tr}\, x^2&=&x_a^{\;\,b}x_b^{\;\,a}=1\ ,\cr &&\cr
{\rm tr}\, x^3&=&x_a^{\;\,b}x_b^{\;\,c}x_c^{\;\,a}=1\ ,\cr&&\cr
.&.&.\cr &&\cr
{\rm tr}\, x^n&=&x_{a_1}^{\;\,a_2}x_{a_2}^{\;\,a_3}\cdots x_{a_n}^{\;\,a_1}=1\ .\label{one1sx}
\eeqa
Since  $[x_a^{\;\,b}]$ is a finite dimensional matrix, there is a finite number of independent such conditions on $x_a^{\;\,b}$.  More specifically, there is a maximum number of $n=(p+q)^2$ independent conditions on the $(p+q+1)\times  (p+q+1)$ on $[x_a^{\;\,b}]$ (excluding  tr$\,x=1$). So, in particular, from tr$\,x=x_a^{\;\,a}=1$ follows that ${\kappa}_a^{\;\,b}$ is traceless, i.e. $\kappa_{p+q+1}^{\;\;\;\,p+q+1}$ is not independent: $\kappa_{p+q+1}^{\;\;\;\,p+q+1}=-{\kappa}_i^{\;\,i}$.

Now we can trivially repeat this construction on the coordinate patch spanned by the local affine coordinates $( \zeta^i,\zeta^*_i)$. Using (\ref{ohptwon}) we have
\beqa
x_i^{\;\,j}=\frac {\zeta^*_i\zeta^j}{{\cal Z}^2}\ , &\ \ & x_{p+q+1}^{\;\;\;\,i}=\;-\frac {\zeta^i}{{\cal Z}^2}\ ,\cr&&\cr  x_i^{\;\,p+q+1}=\;\frac {\zeta^*_i}{{\cal Z}^2}\ , &\ \ & x_{p+q+1}^{\;\;\;\,p+q+1}=\;-\frac {1}{{\cal Z}^2}\ .\label{trefte1}
\eeqa
It is because the  embedding coordinates are in general divergent in the limit~(\ref{thesmptlmt}), that we call this the asymptotic limit. (Components of $x_a^{\;\,b}$ may vanish in the limit in the special cases where $\zeta_i=0$.) The action of the Killing vectors $ \kappa_i^{\;\,j}$ on functions $f$ on the coordinate patch is written exactly as in (\ref{threftwnx})
\be
{ \kappa}_a^{\;\,b} f=-i\{x_a^{\;\,b},f\}\ .\label{threftwn}
\ee
Upon using~(\ref{pbsfrztab}) we can explicitly verify that $ { \kappa}_a^{\;\,b}$ has the form~(\ref{KvsncmtvCP11alc}) (though, of course, this should be obvious from the derivation of (\ref{pbsfrztab}) from (\ref{pbsfzzs})).

For the case of $EAdS_2$, the three real embedding coordinates $x_1,x_2,x_3$ of the section \ref{SectionEAdS2} are recovered from $x_a^{\;\,b}$ by setting
\be
x_1=x_1^{\;\,2}-x_{2}^{\;\,1}\ , \   x_2=-i( x_1^{\;\,2}+x_{2}^{\;\,1})\ , \   x_3= x_1^{\;\,1} -x_{2}^{\;\,2}\ .
\ee
There is only one independent constraint in this case, namely
\be
x_1^2+x_2^2-x_3^2= -2x_a^{\;\,b}x_b^{\;\,a}+(x_a^{\;\,a})^2=-1\ .
\ee

\subsubsection{Canonical coordinates}

Following the previous section, the next step is to perform the Darboux transformation.  As was mentioned above we have not found a single expression for the Darboux transformation that applies for all  $\mathbb{CP}^{p,q}$ spaces.  The difficulty is due to our restriction that the resulting canonical coordinates $(y_i,y_i^*)$ are valid for the whole of $ \mathbb{C}^{p+q}$, in order that  there are no boundaries on our domain in the corresponding quantized theory.  As stated above, we shall give the Darboux transformation for various examples in the sections which follow.  As in the previous case of $EAdS_2$, we find  that the Jacobian of the Darboux transformation goes like
\be
\Big|\frac{\partial(\zeta,\zeta^*)}{\partial(y,y^*)}\Big|={\cal Z}^{2(p+q+1)}\;,\label{jcbynfDbx}
\ee
and hence in terms of  the canonical coordinates, the geometric measure  is proportional to the flat measure
\beqa
d\mu_{\tt geom}(\zeta,\zeta^*)&=&\frac{1}{2^{p+q}}
\, dy^1 \wedge  \cdots\wedge dy^{p+q} \wedge dy^*_1\wedge\cdots \wedge dy^*_{p+q}\ .\label{mugeomyys}
\eeqa

In order to proceed further, we need to assume a Darboux transformation   for
$\mathbb{CP}^{p,q}$ that takes the  local affine coordinates $( \zeta^i,\zeta^*_i)$ to coordinates $(y_i,y_i^*)$ spanning  all of $ \mathbb{C}^{p+q}$ which satisfies the canonical Poisson bracket relations
\be
\{y_i,y_j^*\}=-i\delta_{ij}\;,\quad \{y_i,y_j\}=\{y^*_i,y^*_j\}=0\ .\label{cnclpsnbks}
\ee
for all $i,j=1,...,p+q$. We do not have a general proof of this existence, nor that~(\ref{jcbynfDbx}), and hence~(\ref{mugeomyys}) in general hold, but we are able  to find such transformations for the examples in Sections 4 and 5.

\subsection{Quantization}

Generalizing the procedure that was adapted for $EAdS_2$, we   perform canonical quantization, replacing the   coordinates $(y_i,y^*_i)$ by the set of  operators   $(\hat y_i,\hat y^\dagger_i)$ satisfying commutation relations
\be
[\hat y_i,\hat y^\dagger_j]=\kbar \delta_{ij}\quad ,\quad [\hat y_i,\hat y_j ]=[\hat y^\dagger_i,\hat y^\dagger_j]=0\;,\label{crsfrzzdgr} \ee
$\kbar$ once again being the noncommutative parameter.  This is the algebra for $p+q$ harmonic oscillators.  The lowering and raising operators, $\hat a_i$ and $\hat a^\dagger_i$, are obtained by  rescaling   $ \hat y_i$ and $\hat y^\dagger_i$,  respectively
\beqa
\hat a_i=\frac 1{\sqrt{\kbar}}{\hat y_i} \quad ,\quad \hat a_i^\dagger=\frac 1{\sqrt{\kbar}}{\hat y^\dagger_i}\ .
\eeqa
Then $[\hat a_i,\hat a^\dagger_j]=\delta_{ij}$ and   $[\hat a_i,\hat a_j]= [\hat a^\dagger _i,\hat a^\dagger_j]=0$ for all $i,j=1,...,p+q$.
$\hat a_i$ and $\hat a^\dagger_i$ act on the infinite-dimensional Hilbert space ${\cal H}$, now  spanned by orthonormal  states
\be {| n\rangle}= |n_1,...,n_{p+q}\rangle=\frac{(\hat a^\dagger_1)^{n_1}\cdots (\hat a^\dagger_{p+q})^{n_{p+q}}
}{\sqrt{n_1!\cdots n_{p+q}!}}| 0\rangle\;,\label{stsndsctrp}
\ee
where $n_{i} $ are non-negative integers.  The bottom state $| 0\rangle=|0,...,0\rangle$ is annihilated by any  $\hat a_{i} $,  and has unit norm $\langle 0| 0\rangle=1$.

It is straightforward to generalize the coherent states~(\ref{2dchrntstt}) and {Wick}-Voros star product~(\ref{2dVrsstprd}) to $ \mathbb{C}^{p+q}$.  The former are given by
\be
|\vec \alpha\rangle=|\alpha_1,...,\alpha_{p+q}\rangle =e^{-\frac {|\alpha|^2}2}e^{\alpha_i \hat a_i^\dagger}|\vec 0\rangle\in {\cal H}\;,
\ee
where  $\alpha_i$ are complex eigenvalues  of $\hat a_i$, $\hat a_i |\vec \alpha\rangle=\alpha_i | \vec\alpha\rangle$, and  $|\alpha|^2=\alpha^*_i\alpha_i$.
The completeness relation and normalization condition are now
\beqa
&&\int d\mu(\vec \alpha,\vec\alpha^{\,*})|\vec \alpha\rangle\langle\vec \alpha |=\BI\ ,\cr &&\cr &&
 \langle\vec \alpha|\vec \alpha'\rangle=\exp\Bigl\{{\alpha_i^*{\alpha'_i}-\frac {|\alpha|^2}2-\frac {|\alpha'|^2}2\Bigr\}}\ ,\label{compnorm}
\eeqa
where the  integration measure for coherent states  $d\mu(\vec\alpha,{\vec\alpha}^{\,*})$ is
\be
d\mu(\vec\alpha,{\vec\alpha}^{\,*})=\Bigl(\frac i{2\pi}\Bigl)^{p+q}\,d\alpha_1\wedge d\alpha_1^*\,\wedge\cdots\wedge\, d\alpha_{p+q}\wedge d\alpha_{p+q}^*\ .\label{msrfrchsts}
\ee
Upon doing the rescaling back to canonical coordinates, $y_i={\sqrt{\kbar}}{  \alpha_i}$, we see that it agrees, up to a constant factor,  with the geometric measure~(\ref{mugeomyys}).  Symbols of operators are defined as in~(\ref{2dvrssmbl}), while the {Wick}-Voros product of symbols is
\be
[ {\cal A}\star {\cal B}](\vec y,\vec y^{\,*})={\cal A}(\vec y,\vec y^{\,*})\,\exp
\biggl\{\kbar\sum_{i=1}^{p+q} \overleftarrow{\frac\partial{\partial y_i}}\;\overrightarrow{\frac{ \partial}{\partial y_i^*}}\biggr\}\,{\cal B}(\vec y,\vec y^{\,*})\ .\label{gnrlvrsprd}
\ee
Then the star commutator gives a realization of the fundamental commutaton relations~(\ref{crsfrzzdgr}), and the requirements~(\ref{clmtfstrctr}) for the commutative limit  are satisfied. The star product can be re-expressed in terms of local affine coordinates.   For the examples that follow, as well as the one in section two, we find that the star product reduces to the ordinary product in the asymptotic limit~(\ref{thesmptlmt}).

To define the noncommutative version of $\mathbb{CP}^{p,q}$ we should
construct the noncommutative analogues of the matrix elements   $x_a^{\;\,b}$.  Denoting them by $X_a^{\;\,b}$, we demand  that they  satisfy $su(p+1,q)$ commutation relations
\be
[ X_a^{\;\,b},  X_c^{\;\,d}]_\star=-\kbar(\delta _a^d\, X_c^{\;\,b}- \delta _c^b\, X_a^{\;\,d})\;,\label{ncsu12}
\ee
as well as the analogues of the conditions~(\ref{one1sx}).  The analogues of these conditions fix the Casimirs of the algebra, restricting the allowable representations of   $su(p+1,q)$  of the noncommutative theory.  We, of course, demand that $X_a^{\;\,b}\rightarrow x_a^{\;\,b}$ when $\kbar \rightarrow 0$. In   Sections 4 and 5 we shall provide  perturbative expansions in $\kbar$  for $X_a^{\;\,b}$ as functions of local coordinates for the examples  $\mathbb{CP}^{1,1}$ and  $\mathbb{CP}^{0,2}$, respectively.

Given $X_a^{\;\,b}$ it is then easy to  define  noncommtuative analogues ${ \kappa}_{\,a}^{\star\,b}$ of the Killing vectors.  Generalizing~(\ref{ncdffKcs}) the action of  ${ \kappa}_{\,a}^{\star\,b}$ on functions $f$ on noncommutative $\mathbb{CP}^{p,q}$, we have
\be
\,[{ \kappa}_{\,a}^{\star\,b} f](X)=-\frac 1{\kbar}[X_a^{\;\,b},f]_\star \ . \label{kpuhstrdf}
\ee
Then $ \kappa_{\,a}^{\star\,b}$  are deformations of the Killing vectors $\kappa_a^{\;\,b}$, with the deformation vanishing in the commutative limit $\kbar \rightarrow 0$.   In order to extract the leading order corrections to $  \kappa_a^{\;\,b}$, we need to obtain $[X_a^{\;\,b},f]_\star $ up to second order in $\kbar$.  Even though  $ \kappa_{\,a}^{\star\,b}$ are  deformations of the Killing vectors, they satisfy the same algebra as  $  \kappa_a^{\;\,b}$, namely the $su(p+1,q)$ isometry algebra
\be
[ \kappa_a^{\star\,b}, \kappa_c^{\star\,d}]=\delta _a^d\, \kappa_c^{\star\,b}- \delta _c^b\, \kappa_a^{\star\,d}
\ee
For the two examples which follow, as well as the  one in Section 2, we get that the deformation  of the Killing vectors  vanishes in the asymptotic limit~(\ref{thesmptlmt}).

\section{ $\mathbb{CP}^{1,1}$}
\setcounter{equation}{0}

In this section and the next one we write down the explicit Darboux transformation from local affine coordinates, and perform the quantization procedure as outlined previously.

Here the example is $\mathbb{CP}^{1,1}\simeq H^{2,3}/S^1\simeq  SU(2,1)/U(1,1)$. It can be constructed from  $\mathbb{C}^{2,1}$, spanned by $z^a,\; a=1,2,3$. $\mathbb{CP}^{1,1}$ is then defined by the constraint~(\ref{ess7}), which becomes $|z^1|^2+|z^2|^2-|z^3|^2=1$, along with the equivalence relation~(\ref{eqvrltn}).

There are two complex affine coordinates $\zeta_i$,  $i=1,2$, along with their complex conjugates.   In this case, the background metric on the reduced space is Euclidean, diag$(+,+)$.
The condition~(\ref{wonpt8}) leads to the  restriction  that the local affine coordinates are defined on a real four dimensional space with a solid three-sphere removed, \be {\cal Z}^2=|\zeta_1|^2+|\zeta_2|^2-1\;>\;0\ee  The quantity ${\cal Z}^2$  spans the positive real line, excluding the origin which  corresponds to the asymptotic limit,~(\ref{thesmptlmt}) or  ${\cal Z}^2\rightarrow 0$. While the background metric  for the coordinates is Euclidean, the Fubini-Study metric~(\ref{FSnafncrds}) has a Lorentzian signature. The latter solves the sourceless Einstein equations with $\Lambda=3$~\cite{Stern:2018wud}.

There are eight real embedding coordinates~(\ref{trefte1}), $x_a^{\;\,b}$,  with  tr$\,x=1$.  Since   $\mathbb{CP}^{1,1}$ has four real dimensions, $x_a^{\;\,b}$ are subject to four additional independent conditions~(\ref{one1sx}).

\subsection{Darboux map}

Here we give the transformation  from local affine coordinates to canonical coordinates $(y_i,y_i^*)$, $i=1,2$, satisfying~(\ref{cnclpsnbks}).  As stated previously, we require the domain of the latter to be all of ${\mathbb{C}}^2$, unlike the domain of local affine coordinates.
Up to canonical transformations, the Darboux transformation is given by
\be
y_i=\left\{ \matrix{
	\sqrt{\frac{|\zeta_i|^2}{{\cal Z}^2}-\frac 12}\;\,\frac{\zeta_i}{|\zeta_i|}\;,&\qquad\frac{|\zeta_i|^2}{{\cal Z}^2}>\frac 12 \cr\sqrt{\frac{|\zeta_i|^2}{{\cal Z}^2}-\frac 12}\;\,\frac{\zeta_i^*}{|\zeta_i|}\;,&\qquad\frac{|\zeta_i|^2}{{\cal Z}^2}<\frac 12}\right.
\ .\label{DrbxCp11}
\ee
{Note that the square root is not} {necessarily} {real. To see that the coordinates cover the full complex plane once	 let us express them as:
\beqa
y_1&=&\sqrt{\frac{1}{2{\cal Z}^2}\Big||\zeta_1|^2-|\zeta_2|^2+1\Big|}\times\left\{ \matrix{ \exp{\{i\arg\zeta_1\}},&\ |\zeta_1|^2-|\zeta_2|^2+1>0\cr \exp{\{-i\arg\zeta_1\}},&\ |\zeta_1|^2-|\zeta_2|^2+1<0}\right.\cr&&\cr
\ y_2&=&\sqrt{\frac{1}{2{\cal Z}^2} \Big||\zeta_2|^2-|\zeta_1|^2+1}\Big|\times \left\{ \matrix{ \exp{\{i\arg\zeta_2\}},&\ |\zeta_2|^2-|\zeta_1|^2+1>0\cr \exp{\{-i\arg\zeta_2\}},&\ |\zeta_2|^2-|\zeta_1|^2+1<0}\right.\qquad
\eeqa
One can see that by fixing $\zeta_2$, and letting $\zeta_1$ be  arbitrary, $y_1$ covers the complex plane, and of course the same holds exchanging 1 with 2. The asymptotic limit is
\be
r^2=|y_1|^2+|y_2|^2=\frac1{{\cal Z}^2}\to \infty \ .
\ee
}
The Jacobian of the Darboux  transformation is $\Big|\frac{\partial (\zeta,\zeta^*)}{\partial (y,y^*)}\Big|={\cal Z}^6$ in agreement with~(\ref{jcbynfDbx}), and so we  recover  the flat measure~(\ref{mugeomyys}).

Substituting the Darboux transformation  in the expressions for the   embedding coordinates~(\ref{trefte1}) gives
\beqa
x_i^{\;\,j}=\sqrt{\Bigl(|y_i|^2+\frac 12\Bigr)\Bigl(|y_j|^2+\frac 12\Big)}\,\frac {y^*_iy_j}{|y_i||y_j|}\quad ,&& x_3^{\;\,i}=\;-\sqrt{|y_i|^2+\frac12}\;\frac {ry_i}{|y_i|}\ ,\cr&&\cr  x_i^{\;\,3}=\;\sqrt{|y_i|^2+\frac 12}\;\frac {ry_i^*}{|y_i|}\quad , && x_3^{\;\,3}=\;-r^2\quad ,\label{xijnccs}
\eeqa
$r$ being the positive square root of $r^2$.
We can then check that the constraints~(\ref{one1sx}) and the $su(2,1)$ Poisson bracket algebra~(\ref{psfrtrclssx}) hold.
 Substituting~(\ref{xijnccs}) into~(\ref{threftwn})  gives the Killing vectors   in terms of canonical coordinates.

\subsection{Quantization}

Quantization proceeds as in Section 3, with the Hilbert space  ${\cal H}$ being that of a two-dimensional harmonic oscillator. The {Wick}-Voros star product is given in~(\ref{gnrlvrsprd}), and   can be re-expressed in terms of local affine coordinates by making the replacement
\beqa\frac \partial{\partial y_1}&\rightarrow&\frac{{\cal Z}}{2 \sqrt{2}\, {\zeta_1}}\Biggl\{- |{\zeta_1}| \sqrt{
|{\zeta_1}|^2-|{\zeta_2}|^2+1}
\left({\zeta^*_2}\frac\partial{\partial \zeta^*_2}+{\zeta_2}\frac\partial{\partial \zeta_2}\right)\cr&&\cr
&& \;+\;\frac{-|{\zeta_2}|^2+1}{|{\zeta_1}| \sqrt{
|{\zeta_1}|^2-|{\zeta_2}|^2+1}}\Bigl({\zeta^*_1}\frac\partial{\partial \zeta^*_1}
  +{\zeta_1}
  \frac\partial{\partial \zeta_1}\Bigr) \,+\,\frac{|{\zeta_1}|  (|{\zeta_1}|^2
  -|{\zeta_2}|^2 +2)}{ \sqrt{
|{\zeta_1}|^2-|{\zeta_2}|^2+1}}\; \kappa_1^{\;\,1}\Biggr\}\;,\;\quad\qquad\label{dydywon}\eeqa
along with the corresponding replacement for $\frac \partial{\partial y_2}$, obtained by switching the coordinate indices $1$ and $2$ in~(\ref{dydywon}).  Since they both contain the over-all  factor of ${\cal Z}$, it follows that  the star product reduces to the ordinary product in the asymptotic limit, ${\cal Z}\rightarrow 0$.

Next we construct the noncommutative analogues $ X_a^{\;\,b}$ of the embedding coordinates~(\ref{xijnccs}).
We take the following ans\"atse
\be  [ X_a^{\;\,b}]=\pmatrix{|y_1|^2+\frac 12&\quad {\cal R}_1 \frac{y_1^*}{|y_1|}\star{\cal R}_2 \frac{y_2}{|y_2|}&\quad {\cal R}_1 \frac{y_1^*}{|y_1|}\star{\cal S} \cr {\cal R}_2 \frac{y_2^*}{|y_2|}\star {\cal R}_1 \frac{y_1}{|y_1|}  & |y_2|^2+\frac 12 &\quad {\cal R}_2 \frac{y_2^*}{|y_2|}\star{\cal S} \cr -{\cal S}\star {\cal R}_1 \frac{y_1}{|y_1|}  &-{\cal S}\star{\cal R}_2 \frac{y_2}{|y_2|}  &-r^2} \;,\label{ncfrateefv11}\ee
where we assume that  ${\cal R}_i$  is a real function of $|y_i|^2$, and ${\cal S}$ is a real function of $r^2$.

In order to recover~(\ref{xijnccs}) in the commutative limit, we need that ${\cal R}_i \rightarrow{\cal R}_i^{(0)}=\sqrt{|y_i|^2+\frac 12}$, and  ${\cal S}\rightarrow{\cal S}^{(0)}=r$ when $\kbar \rightarrow 0$.  Away from the commutative limit, ${\cal R}_i $ and
 ${\cal S}$ can be obtained {as a perturbative expansion is $\kbar$}
\beqa
 && {\cal R}_i\;=\;{\cal R}^{(0)}_i \;+\;\kbar{\cal R}^{(1)}_i \;+\;\kbar^2 {\cal R}^{(2)}_i\;+\;{\cal O}(\kbar^3)\ ,\cr&&\cr
&&{\cal S}\;=\;{\cal S}^{(0)}\;+\;\kbar {\cal S}^{(1)}\;+\;\kbar^2 {\cal S}^{(2)}\;+\;{\cal O}(\kbar^3)\ .
\label{nccs2X3S}
\eeqa
For this we require that $ X_a^{\;\,b}$  satisfy the $su(2,1)$ star commutator algebra~(\ref{ncsu12}).
For the leading two corrections we find
\beqa
{\cal R}^{(1)}_i&=&-\frac 1{32\,|y_i|^2\Bigl(|y_i|^2+\frac 12\Bigr)^{3/2}}+\frac{c_1}{8\sqrt{|y_i|^2+\frac 12}}\quad ,
\cr &&\cr
 {\cal R}^{(2)}_i&=&-\frac{7+48|y_i|^2+128|y_i|^4}{2048 |y_i|^4\Bigl(|y_i|^2+\frac 12\Bigr)^{7/2}}-\frac{3c_1}{128\Bigl(|y_i|^2+\frac 12\Bigr)^{5/2}}
\cr&&\cr  &&   \quad - \frac{c_1+c_1^2|y_i|^2}{128 |y_i|^2\Bigl(|y_i|^2+\frac 12\Bigr)^{3/2}}+\frac{c_2}{8\sqrt{|y_i|^2+\frac 12}}
\eeqa
and
\be\;\; {\cal S}^{(1)}\;=\;-\frac {1+c_1}{8r}\quad ,\quad {\cal S}^{(2)}\;=\;-\frac{c_1^2+6c_1+7}{128 r^3}-\frac {c_2}{8r}\;,\quad \qquad\qquad \ee
where $c_1$ and $c_2$ are arbitrary real constants.   While tr$\,X= X_a^{\;\,a}=1$, as in the commutative theory, there are noncommutative corrections to the constraints~(\ref{one1sx}).  For example,
\beqa
{\rm tr}\, X^2&=&X_a^{\;\,b}\star X_b^{\;\,a}=1+(c_1+2)\kbar +\Bigl( c_2+\frac 32 c_1+\frac 3 8 c_1^2\Bigr)\kbar^2+{\cal O}(\kbar^3)\ ,\cr&&\cr
 {\rm tr}\, X^3&=&X_a^{\;\,b}\star X_b^{\;\,c}\star X_c^{\;\,a}=1+\Bigl(\frac 32 c_1+4\Bigr)\kbar+\frac 32\Bigl( c_2+3 c_1+\frac 12 c_1^2 +\frac 83\Bigr)\kbar^2\cr&&\cr
 &&+{\cal O}(\kbar^3)\ .\cr&&
\eeqa
They correspond to the quadratic and cubic Casimir operators for $su(2,1)$.
We note that there  is no choice of $c_1$ and $c_2$ for which the noncommutative corrections in both  tr$X^2$ and   tr$X^3 $ disappear.

Upon writing the result for the expansion~(\ref{nccs2X3S}) in terms of local affine coordinates one gets
\beqa
\frac{y_i}{|y_i|} {\cal R}_i&=&\frac{\zeta_i}{{\cal Z}}\Biggl\{1 \;+\;\frac\kbar{16|\zeta_i|^4}\,\frac{{\cal Z}^6}{({\cal Z}^2-2|\zeta_i|^2)}\cr&&\cr&&\;\;-\;\frac{\kbar^2}{512}\;\frac{{\cal Z}^8}{|\zeta_i|^8}\;\frac{(63|\zeta_i|^4-50|\zeta_i|^2({\cal Z}^2-|\zeta_i|^2)+15({\cal Z}^2-|\zeta_i|^2)^2}{({\cal Z}^2-2|\zeta_i|^2)^2}\;+\;{\cal O}(\kbar^3)\Biggr\}\cr&&\cr &&\cr
{\cal S}&=&\frac{1}{{\cal Z}}\Biggl\{1\;-\;\frac{\kbar}8 \,{\cal Z}^2\;-\;\frac{7\kbar^2}{128}\,{\cal Z}^4 \;+\;{\cal O}(\kbar^3)\Biggr\}\;,\cr&& \eeqa
where for simplicity we set $c_1=c_2=0$.
The zeroth order terms in $\kbar$ correspond to the commutative result.  When substituted into~(\ref{ncfrateefv11}), and extracting the zeroth order terms, we recover the formulae~(\ref{trefte1}) for embedding coordinates.
The noncommutative corrections to $\frac{y_i}{|y_i|} {\cal R}_i$ are not valid near $\zeta_i=0$.
The noncommutative corrections to $\frac{y_i}{|y_i|} {\cal R}_i$ and ${\cal S}$, and hence $X_a^{\;\,b}$,  contain factors of ${\cal Z}$, and so, away from  $\zeta_i=0$, these corrections vanish in the asymptotic limit ${\cal Z}\rightarrow 0$.   For this we also use the above result that the star product, when expressed in terms of local affine coordinates,  reduces to the ordinary product in the asymptotic limit. Finally we can construct the  series expansion for the noncommutative analogue ${ \kappa}_{\,a}^{\star\,b}$ of the Killing vector on  $CP^{1,1}$ using~(\ref{kpuhstrdf}).
The above arguments show  that  they too  reduce to the
commutative Killing vectors~(\ref{KvsncmtvCP11alc}) in the asymptotic limit.

\section{$\mathbb{CP}^{0,2}$}
\setcounter{equation}{0}

Like $\mathbb{CP}^{1,1}$, $\mathbb{CP}^{0,2}$ has four real dimensions. $\mathbb{CP}^{0,2}\simeq H^{4,1}/S^1\simeq SU(2,1)/U(2)$ can be built from $\mathbb{C}^{1,2}$, spanned by $z^a,\;a=1,2,3$, using the the constraint~(\ref{ess7}), which now becomes  $|z^1|^2-|z^2|^2-|z^3|^2=1$,
along with the equivalence relation~(\ref{eqvrltn}). This means that $|z^1|\ge 1$, and also that $|z^1|>|z^2|$ or $|z^3|$.

Once again there are two complex affine coordinates $\zeta_i$, $i=1,2$, along with their complex conjugates. They are defined by $ \zeta^i=\frac{z^i}{z^{3}}$, $z^3\ne 0$. Unlike the case with $\mathbb{CP}^{1,1}$, here the indices  $i,j,...$ are raised and lowered with the Lorentzian metric, diag$(+,-)$.  So here~(\ref{wonpt8}) implies that
\be
{\cal Z}^2=|\zeta_1|^2-|\zeta_2|^2-1\;>\;0
\ee
and so $|\zeta_1|> 1$. This  restriction means  that the local affine coordinates are defined on a real four dimensional space with a solid three-hyperboloid removed. The boundary of this region once again corresponds to the asymptotic limit~(\ref{thesmptlmt}), ${\cal Z}^2\rightarrow 0$.  While the background metric is Lorentzian, the Fubini-Study metric~(\ref{FSnafncrds}) for $\mathbb{CP}^{0,2}$ has a Euclidean signature. This is opposite the  situation with $\mathbb{CP}^{1,1}$. As with $\mathbb{CP}^{1,1}$, the Fubini-Study metric solves the sourceless Einstein equations with $\Lambda=3$~\cite{Stern:2018wud}.

\subsection{Darboux map}

We now give the  transformation from the local affine coordinates  $( \zeta^i,\zeta^*_i), \; i=1,2$, to canonical coordinates $(y_i,y_i^*)$, satisfying Poisson brackets~(\ref{cnclpsnbks}).  We note that the indices for the former are raised and lowered using the Lorentzian metric, but the latter coordinates are  defined on a two-dimensional  complex Euclidean space.  Because of this  fact it is helpful to perform an intermediate step.  For this we  recognize that  local affine coordinates are not unique.  Instead of using the coordinates
$( \zeta^i,\zeta^*_i)$,  as defined in~(\ref{ohptwon}),
we can choose to work with the alternative set of  coordinates  $( \xi^n,\xi^*_n),\; n=1,2$, where $ \xi^n=\frac{z^{n+1}}{z^{1}}$,  $ z^1\ne 0$. In contrast with $( \zeta^i,\zeta^*_i)$, for these coordinates, the indices  $n,m,...$ are raised and lowered with the {\it Euclidean} metric, diag$(-,-)$.  The transformation between the two sets of local affine coordinates (in the overlapping region) is therefore something like a Wick rotation of the parameter space,  although the signature of the Fubini-Study metric, of course, remains Euclidean.  The  transformation between the two sets of local affine coordinates is given by
\be
\xi^1=\frac{\zeta^2}{\zeta^1}\qquad\quad\xi^2=\frac{1}{\zeta^1}\;,\qquad\quad \zeta^1,\xi^2\ne0 \ .\label{frsvnwn}
\ee
The two sets of coordinates are valid on different domains and the transformation applies in the overlapping region. From~(\ref{frsvnwn})
\be
1- |\xi_1|^2- |\xi_2|^2 =\frac{{\cal Z}^2}{|\zeta_1|^2}=\frac 1{|z^1|^2}> 0\;,
\ee
and hence    $( \xi^n,\xi^*_n)$ span the interior of a three-sphere of radius one, $|\xi_1|^2+ |\xi_2|^2< 1$.  As usual the boundary corresponds to the asymptotic limit $|\xi_1|^2+ |\xi_2|^2\rightarrow 1$. The Fubini-Study metric and Poisson brackets can be re-expressed in terms of the new local affine coordinates  $( \xi^n,\xi^*_n)$.

It is now not difficult to find the map from the affine coordinates  $( \xi^n,\xi^*_n)$ to canonical coordinates $(y^i,y_i^*)$, $i=1,2$, having the desired properties.  Up to canonical transformations, it  is
\be
y_1=\frac {i\,\xi^*_1}{\sqrt{1- |\xi_1|^2- |\xi_2|^2}}\quad ,\quad y_2=\frac {-i\,\xi^*_2}{\sqrt{1- |\xi_1|^2- |\xi_2|^2}} \ . \label{Drbufrmxi}
\ee
There are no restrictions on the domain of  $(y_i,y_i^*)$, i.e., they span all of $\mathbb{C}^{2}$.  To see this note that
\be
r^2=\frac {|\xi_1|^2+ |\xi_2|^2}{1- |\xi_1|^2- |\xi_2|^2}\ge 0\;,\label{eekfrsq}
\ee
where we once again define $ r^2= |y_1|^2+|y_2|^2$.  The right hand side of~(\ref{eekfrsq}) spans the entire positive real line. Moreover, $|y_1|^2$ and  $|y_2|^2$  span the entire positive real line.  Just as with the case of   $\mathbb{CP}^{1,1}$, $r^2\rightarrow \infty$ is the boundary limit.

Using~(\ref{frsvnwn}) and~(\ref{Drbufrmxi}), we can write the Darboux map from the original set of affine coordinates  $(\zeta^i,\zeta^*_i)$. It  is
\be
y_1=\frac{-i\zeta^*_2}{{\cal Z}}\sqrt{\frac{\zeta_1}{\zeta^*_1}}\quad ,\qquad
y_2=\frac{i}{{\cal Z}}\sqrt{\frac{\zeta_1}{\zeta^*_1}}\ .\label{dbxnpcp02}
\ee
{This is an extension of the Darboux map for  $EAdS^2$ (\ref{2dDbxtrn}), where $\zeta$ and $y$ now correspond to $\zeta_1$ and $-iy_2$, respectively.}
The Jacobian of the transformation is $\Big|\frac{\partial (\zeta,\zeta^*)}{\partial (y,y^*)}\Big|={\cal Z}^6$, so we again recover the flat geometric measure when expressed in terms of  canonical coordinates.

Writing  the embedding coordinates~(\ref{trefte1}) in terms of  canonical coordinates gives
\be
[ x_a^{\;\,b}]=\pmatrix{r^2+1&\quad   i{y^*_1}\sqrt{r^2+1}&\quad  i{y^*_2}\,\sqrt{r^2+1}\cr   i{y_1}\,\sqrt{r^2+1}& -|y_1|^2 &- y_2^*{y_1}\cr   iy_2\,\sqrt{r^2+1} & - {y^*_1}y_2&-|y_2|^2}  \quad .\label{frateefv}
\ee
We can then check that the constraints~(\ref{one1sx}) and the $su(1,2)$ Poisson bracket algebra~(\ref{psfrtrclssx}) hold.
 Substituting~(\ref{frateefv}) into~(\ref{threftwn})  gives the Killing vectors   in terms of canonical coordinates.

\subsection{Quantization}

Quantization proceeds as in the previous section.  The algebra of observables is again that  of  a two-dimensional harmonic oscillator, which is realized with the {Wick}-Voros star product~(\ref{gnrlvrsprd}).
The star product  can again be re-expressed in terms of the original local affine coordinates  $( \zeta^i,\zeta^*_i)$, now by making the replacement
\beqa
\frac \partial{\partial y_1}&\rightarrow&\frac{i {\cal Z}}{2\zeta_1|\zeta_1|}\Biggl\{
\zeta_2\biggl(\zeta_1\,\frac\partial{\partial \zeta_1}+\zeta_1^*\,\frac\partial{\partial \zeta^*_1}\biggr)+2|\zeta_1|^2\,\frac\partial{\partial \zeta^*_2}\Biggr\}\ ,\cr&&\cr
\frac \partial{\partial y_2}&\rightarrow&\frac{-i {\cal Z}}{2\zeta_1|\zeta_1|}\Biggl\{\zeta_1\,\frac\partial{\partial \zeta_1}+(1-2{|\zeta_1}|^2)\zeta^*_1\,\frac\partial{\partial \zeta^*_1}-2{|\zeta_1}|^2\zeta^*_2\,\frac\partial{\partial \zeta^*_2}
\Biggr\}\ .
\eeqa
Because of the over-all  factor of ${\cal Z}$, it follows that  the star product reduces to the ordinary product in the asymptotic limit, ${\cal Z}\rightarrow 0$.

Next we construct the noncommutative analogues $ X_a^{\;\,b}$ of the embedding coordinates~(\ref{frateefv}).
We try writing
\be
[ X_a^{\;\,b}]=\pmatrix{r^2+1&\quad   i{y^*_1}\star{\cal S}&\quad  i{y^*_2}\star{\cal S}\cr   i{\cal S}\star{y_1}& -|y_1|^2 &- y_2^*{y_1}\cr   i{\cal S}\star y_2 & - {y^*_1}y_2&-|y_2|^2} \ ,\label{ncfrateefv}
\ee
where we assume that ${\cal S}$ is a real function of $r^2$.
We need that  ${\cal S}\rightarrow{\cal S}_0=\sqrt{r^2+1} $ when $\kbar \rightarrow 0$, in order to recover~(\ref{frateefv}) in the commutative limit. In order to obtain ${\cal S}$ away from the commutative limit, we require that $ X_a^{\;\,b}$  satisfy the $su(1,2)$ star commutator algebra~(\ref{ncsu12}). We can then get ${\cal S}$  in a perturbative expansion in $\kbar$.  So as before we write ${\cal S}={\cal S}_0+\kbar {\cal S}_1+\kbar^2 {\cal S}_2+\cdots$.  For the leading two corrections we get
\be
{\cal S}_1=-\frac{r^2}{8(r^2+1)^{3/2}}\;,\qquad\quad {\cal S}_2=\frac{r^2(8-7 r^2)}{128(r^2+1)^{7/2}}\ .
\ee
Once again, while tr$\,X= X_a^{\;\,a}=1$, as in the commutative theory, there are noncommutative corrections to the constraints~(\ref{one1sx}).  For example,
\beqa
{\rm tr}\, X^2&=&X_a^{\;\,b}\star X_b^{\;\,a}=1-2\kbar +{\cal O}(\kbar^3)\ ,\cr&&\cr
{\rm tr}\, X^3&=&X_a^{\;\,b}\star X_b^{\;\,c}\star X_c^{\;\,a}=1-2\kbar -2\kbar^2+{\cal O}(\kbar^3)\ .
\eeqa
In comparing  the expansion found here with the one found for  $\mathbb{CP}^{1,1}$, we note that the latter was expressed in terms of  undetermined integration constants  $c_1$ and $c_2$.  Integration constants may appear for $\mathbb{CP}^{0,2}$ as well upon generalizing the ansatz~(\ref{ncfrateefv}).

From~(\ref{ncfrateefv}), noncommutative corrections to  the embedding coordinates only appear for $X_1^{\;\,2}$,  $X_1^{\;\,3}$,  $X_2^{\;\,1}$ and $  X_3^{\;\,1}$.  After writing the leading order terms for these four matrix elements in the original affine coordinates   $( \zeta^i,\zeta^*_i)$,  we get
\beqa
X_a^{\;\,b}= x_a^{\;\,b}\,\biggl(1-\frac {{\cal Z}^2(1+|\zeta_2|^2)}{8|\zeta_1|^4}\kbar +
\;\frac {{\cal Z}^4(1+|\zeta_2|^2)(8|\zeta_1|^2-15 |\zeta_2|^2-15)}{128|\zeta_1|^8}\kbar^2\cr&&\cr
+\;{\cal O}(\kbar^3)\biggr)\;,
\eeqa
where again this only applies for $(a,b)= (1,2), (1,3), (2,1), (3,1)$. We  find that the corrections contain factors of ${\cal Z}^2$, and so they vanish in the asymptotic limit, ${\cal Z}^2\rightarrow 0$. Finally, we can obtain the leading corrections to the Killing vectors, specifically $\kappa_1^{\;\,2}$,  $\kappa_1^{\;\,3}$,  $\kappa_2^{\;\,1}$ and $  \kappa_3^{\;\,1}$, using the definition~(\ref{kpuhstrdf})  for their noncommutative analogue.  Since they involve taking a star product, which reduces to the ordinary product in the commutative limit, we once again see that all  noncommutative corrections to the Killing vectors vanish in the asymptotic limit.

\section{Concluding remarks}
\setcounter{equation}{0}

In this article we have shown how to perform a unique quantization of $\mathbb{CP}^{p,q}$ which preserves the full $su(p+1,q)$ isometry algebra. For the specific examples considered here we found that noncommutativity is effectively restricted to a limited neighborhood of some origin, and that these quantum spaces approach $\mathbb{CP}^{p,q}$ in the asymptotic limit. It is likely that this is a universal result that applies for all $\mathbb{CP}^{p,q},\; q\ge 1$  quantized in a isometry preserving manner. Just as a strong-weak duality is postulated to exist between gravity on asymptotically $AdS$ spaces and a CFT on the boundary, it is tempting to speculate that a similar duality could exist between gravity on asymptotically  $\mathbb{CP}^{p,q}$ spaces and some boundary field theory.  Adapting the standard techniques to this case, it should be possible to  compute $n-$point correlation  on the boundary, which are  expected  to be consistent with the  $su(p+1,q)$  algebra, rather than the full conformal algebra.  So then if we have that noncommutative $\mathbb{CP}^{p,q}$ is asymptotically $\mathbb{CP}^{p,q}$, there could exist a dual  $SU(p+1,q)$ invariant boundary theory.

As was stated in the text, the main reason we do not have an explicit construction for all quantized   $\mathbb{CP}^{p,q},\; q\ge 1$, and cannot prove asymptotic commutativity in general, is that we do not have a universal construction of the Darboux map.  The Darboux map from local affine coordinates needed to satisfy three requirements, one of which  was that the resulting canonical coordinates cover the entire complex plane.   We found explicit constructions of the map for all examples in two and four dimensions.
Straightforward higher dimensional generalizations of these constructions exist, but they cannot be applied to all cases.
There are two types of higher dimensional extensions:
1)   $\mathbb{CP}^{p,1}$ and 2)  $\mathbb{CP}^{0,q}$.

\begin{enumerate}
\item
 $\mathbb{CP}^{p,1}$, the coordinate patch spanned by the local affine coordinates $( \zeta^i,\zeta^*_i)$ is  ${\mathbb{C}}^{p+1}$ with the region  $|\zeta_1|^2+|\zeta_2|^2+\cdots +|\zeta_{p+1}|^2\le 1$ removed.  The Darboux transformation  to canonical coordinates $(y_i,y_i^*)$, $i=1,2,...,p+1$, can again be given by (\ref{DrbxCp11}).  The latter are defined  on all of ${\mathbb{C}}^{p+1}$.  The expressions for the $su(p+1,1)$ embedding coordinates $x_{ab}$ have the form (\ref{xijnccs}), and their quantum corrections can be computed as in section 4.

\item $\mathbb{CP}^{0,q}$, the coordinate patch spanned by the local affine coordinates $( \zeta^i,\zeta^*_i)$ is  ${\mathbb{C}}^{1,q-1}$ with the region  $|\zeta_1|^2-|\zeta_2|^2-\cdots -|\zeta_{q}|^2\le 1 $ removed.   A Darboux transformation  to canonical coordinates $(y_i,y_i^*)$, $i=1,2,...,q$, is
\be
y_1=\frac{-i\zeta^*_2}{{\cal Z}}\sqrt{\frac{\zeta_1}{\zeta^*_1}}
\;,\quad\cdots \quad ,\quad y_{q-1}=\frac{-i\zeta^*_q}{{\cal Z}}\sqrt{\frac{\zeta_1}{\zeta^*_1}}\;,\qquad\quad y_q=\frac{i}{{\cal Z}}\sqrt{\frac{\zeta_1}{\zeta^*_1}}
\;,\ee
which generalizes (\ref{dbxnpcp02}). The coordinates  $(y_i,y_i^*)$ span  all of ${\mathbb{C}}^{q}$. The expressions for the $su(1,q)$ embedding coordinates $x_{ab}$ become
\be  [ x_a^{\;\,b}]=\pmatrix{r^2+1&\quad   i{y^*_1}\sqrt{r^2+1}&\quad  i{y^*_2}\,\sqrt{r^2+1}&\cdots&i{y^*_q}\,\sqrt{r^2+1} \cr   i{y_1}\,\sqrt{r^2+1}& -|y_1|^2 &- y_2^*{y_1}&\cdots&- y_q^*{y_1}\cr iy_2\,\sqrt{r^2+1} & - {y^*_1}y_2&-|y_2|^2&\cdots&- y_q^*{y_2}\cr  \cdots&\cdots&\cdots&\cdots&\cdots\cr
	iy_q\,\sqrt{r^2+1} & - {y^*_1}y_q&-y_2^* y_q&\cdots&- |y_q|^2}  \;,\ee
generalizing (\ref{frateefv}), while their quantum corrections can be computed as in section 5.
\end{enumerate}

More work is required to obtain the Darboux map for other cases, as it appears that a universal formula does not apply.  One case, in particular, that is not included in 1) and 2), and may be worth pursuing is $\mathbb{CP}^{1,2}$, as it contains Euclidean $AdS_4$ as a submanifold, and its noncommutative version is of possible interest for quantum cosmology.\cite{Steinacker:2019fcb} The noncommutative  analogue of Euclidean $AdS_4$ is constructed from quantized  $\mathbb{CP}^{1,2}$.  Therefore if, as expected, quantized $\mathbb{CP}^{1,2}$ is asymptotically commutative, it should naturally follow that noncommutative $AdS_4 $ is asymptotically anti-de Sitter, having a dual three-dimensional  conformal theory at the boundary.

\end{document}